\documentstyle[12pt,openbib,worldsci]{article}

\def\Im{\mathop{{\cal I}\!m}}

\begin{document}

\font\fortssbx=cmssbx10 scaled \magstep2
%
\hskip.5in \raise.1in\hbox{\fortssbx University of Wisconsin - Madison}
\hfill\parbox{1.25in}{{\bf MAD/PH/784}\\August 1993}
\vspace{1cm}

\begin{center}
{\large\bf The Pomeron as Massive Gluons\footnotemark}\\[.4cm]
M.~B.~Gay~Ducati$^{1,2}$\\[.2cm]
\it $^1$Deparment of Physics, University of Wisconsin, Madison, WI
53706\\[.2cm]
$^2$Instituto de F\'{\i}sica, Universidade Federal do Rio Grande do Sul\\
Av. Bento Gon\c{c}alves, 9500, 91501-970, Porto Alegre, RS, Brazil\\
\end{center}

\footnotetext{{\it Talk presented at the Vth Blois Workshop on Elastic and
Diffractive Scattering}, June 1993.}

\thispagestyle{empty}
\vspace{0.2cm}

\begin{abstract}
  A QCD-Pomeron composed by two
non-perturbative gluons with a dynamically generated  mass, is constructed in
a gauge invariant way. The gluon propagator is infrared-finite. The model
properly describes data on elastic scattering, exclusive $\rho$ production
in deep inelastic scattering (DIS) and the $J/\Psi$-nucleon total cross-section
in terms of a single gluon mass $m_g\simeq0.37$~GeV. The total cross sections
of hadrons with small radii, such as $J/\Psi$, are very sensitive on the
effective gluon mass.
\end{abstract}

\section{Introduction.}

  The Pomeron exchange describes well the diffractive scattering. Within
 QCD the Pomeron is considered as the exchange of two (or more)
gluons~\cite{low,hal}. However, two perturbative gluons cannot
reproduce the experimental results, as can be done with the
exchange of two non-perturbative gluons (NPG)~\cite{cor}. Relating their
 properties
with the QCD vacuum~\cite{lan}, the NPG have a finite correlation length,
or a mass, associated to the gluon
field, which can be understood in terms of gluon
condensates (LN model).

 We proposed a QCD Pomeron model using a solution of the
Schwinger-Dyson equation for the gluon propagator, which contains a dynamically
generated gluon mass~\cite{hal2,gay}. This solution is obtained in a gauge
invariant
way and  it is finite at
$k^2=0$.

 This model describes the data on $\rho$ meson
production in DIS~\cite{aub}, which is claimed as a good test
for the non-perturbative Pomeron~\cite{don}, and the $J/\Psi$-nucleon total
cross
section, in terms of the same gluon mass derived from elastic $pp$ scattering.
  Also, the $J/\Psi$-nucleon total cross-section~\cite{dol}
presents a strong dependence on the dynamical gluon mass.

\section{The NPG exchange model}

In the LN model, Pomeron exchange between quarks has a structure similar to a
photon exchange diagram with amplitude
$i\beta_{0}^{2}(\overline{u}\gamma_{\mu}u)(\overline{u}\gamma^{\mu}u)$,
 where $\beta_{0}$ is the strength of the Pomeron coupling to quarks
\begin{equation}
\beta_{0}^{2}=\frac{1}{36\pi^{2}}\int_{}^{}\,d^2k
\left[g^{2}D(k^{2})\right]^2,
\label{e1}
\end{equation}
where $g^2/4\pi$ is the strength of the non-perturbative coupling
($\beta_{0}^{2}=4\rm~GeV^{-2}$)~\cite{hal2}.
 For an
infrared-finite propagator the integral in Eq.~(\ref{e1}) converges,
as required to reproduce the additive quark rule for
total cross sections~\cite{lan}.

Cornwall~\cite{cor} associates to the gluon a dynamically generated mass
by means of a gauge invariant truncation of the gluonic Schwinger-Dyson
equation.  This solution is given by
$D_{\mu\nu}=-ig_{\mu\nu}D(k^2)$, in the Feynman gauge, and~\cite{cor}
\begin{equation}
D^{-1}(k^2) = \left[k^2+m^2(k^2)\right]bg^2 \ln \left[ \frac{k^2+4m^2(k^2)}
{\Lambda^2} \right],
\label{e2}
\end{equation}
with the momentum-dependent dynamical mass given by
$m^2(k^2) =\break
 m_g^2 \left[
{\ln\left(\frac{k^2+4m_g^2}{\Lambda^2}\right)}/
{\ln\frac{4m_g^2}{\Lambda^2}}\right]^{-12/11}$\llap.
%
 In these expressions $m_g$ is the gluon mass, and $b=(33-2n_f)/48\pi^2$ is
the
leading order coefficient of the $\beta$ function of the renormalization group
equation, where $n_f$ is the number of flavors taken as 3.
The effect of fermion loops  is included in $b$. Formally, $g^2D(k^2)$ is
 independent of the coupling $g$, which is frozen and
 required to be in the range
1.5--2. This constraint and
the value of the Pomeron coupling to quarks, see Eq.~(1), fixes the
non-perturbative propagator~\cite{cor}.

 In Fig.~1 we show the determination
of $\beta_{0}^{2}$ as a function of the gluon mass using Eq.~(\ref{e2}) for
three different values of $\Lambda$. For $\Lambda=300$~MeV and $m_g=380$~MeV we
obtain $\beta_{0}\simeq 2\rm~GeV^{-1}$. This value of the gluon mass is
consistent with $m_g=370$~MeV obtained from fits to the $pp$ total
cross-section, shown in Fig.~2. We find that a gluon mass $m_g=1.2$--$2
\Lambda_{\rm QCD}$ is in
agreement with experiment. We choose $\Lambda=0.3$~GeV, and we obtain that a
variety
of experimental data requires a gluon mass $m_g\simeq{0.37}$~GeV.

\section{Exclusive $\rho$ production in DIS.}

 The exclusive DIS
process $\gamma^*p\rightarrow{\rho}p$ can measure the
Pomeron-quark coupling off-shellness~\cite{don}. The amplitude for
this process behaves as
\begin{equation}
A  \propto\int d{k^2}
\frac{-4k^2+t}{(m_{\rho}^2
+Q^2-t)(-4k^2+Q^2+m_{\rho}^2)}
\left[4\pi\alpha_nD(k^2+t/4)
\right]^2,
\label{e3}
\end{equation}
where $\alpha_n$=$g^2/4\pi$, $q$, $P$ and $P'$ are the four-momenta of the
photon,
incoming and outgoing proton, $Q^2=-q^2$, $t=(P-P')^2$
and $m_{\rho}$
is the $\rho$ meson mass.
The QCD motivated propagator of Eq.~(2) is valid for the entire range of
momentu
m with
$g=1.5$.  We will assume
$g^2(k^2)\simeq g^2/(1+bg^2ln\frac{k^2+m_g^2}{\Lambda^2})$, i.e.\ for
$k^2\rightarrow0$, $g^2(k^2)$ approaches the frozen value of $g$, and
the perturbative coupling is recovered at high momentum.

The differential cross section is given by
\begin{equation}
\frac{d\sigma}{dt}=\left[ \frac{\alpha_{\it{elm}}}{4w^4}|A|^2
{\Phi}^2 \right]Z^2\left[{3F_1(t)}\right]^2,
\label{e4}
\end{equation}
where $\alpha_{\it{elm}}$ is the electromagnetic coupling constant,
$\Phi$ gives the strength of the
$q \overline{q}\rho$
vertex, $F_1(t)$ is the proton elastic form factor, and
$Z=\left(w^2/w_0^2\right)^{0.08+{\alpha}'t}$,
 where $w^2=(P+q)^2$, $w_0^2\simeq{1/\alpha '}$ and
 $\alpha ' =(2~GeV)^{-2}$. This factor
introduces the Pomeron exchange dependence on energy and the
two-gluon exchange
reproduces a Pomeron trajectory $1+0t$. Since at $t=0$ the energy
dependence is very small we can compare the value of $\beta_0$
with the experimental one.

The total cross section for $\gamma^*p\rightarrow{\rho}p$, which is the sum of
the transverse and longitudinal parts $\sigma_{total}=\sigma_T+{\epsilon}
\sigma_L$, is shown in Fig.~3 for $\Lambda=0.3$~GeV, $\langle w \rangle
=12$~GeV and $\epsilon=0.85$. We conclude that $m_g=0.37$~GeV describes the
data~\cite{aub}. There is a strong variation of the cross section with the
gluon mass and that, once $\Lambda$ is fixed, this is a  one-parameter
fit. Some disagreement is expected as we move towards low values of $Q^2$, due
to non-perturbative effects.

\section{$J/\Psi$ - nucleon scattering.}

The amplitude for meson-nucleon scattering in the NPG model
is given by~\cite{dol}
\begin{eqnarray}
A & = & i\frac{32}{9}s\alpha^2\int_{}^{}\,d^2kD(k^2)
D((2{\bf Q}-{\bf k})^2)
2\left[f_M(Q^2)-f_M(({\bf Q}-{\bf k})^2)\right]\nonumber\\
 & &\times 3\left[f_N(Q^2)-f_N
(Q^2 -\frac{3}{2}{\bf Q\cdot k}+\frac{3}{4}k^2)\right],
\label{e5}
\end{eqnarray}
where $s$ is the square of the center of mass energy and $f_M$ and $f_N$
 are respectively the meson and nucleon form factors. The
total cross section is related to this amplitude by
%
$\sigma_T=\Im A(s,t=0)/s$,
%
 and we use form factors in the pole approximation in the calculation. In our
picture of the Pomeron, the relation between cross section and effective radii
of hadrons~\cite{pov} modifies for heavy mesons as $J/\Psi$, since it strongly
depends on the gluon mass.

 In our calculation we compute the coefficient of
the term $s^{0.08}$ in the total cross section~\cite{don} and not the growth
with energy. Then,
for the ratio of cross sections  $\sigma_{\Psi p}/\sigma_{\pi p}$
 we expect a factor of approximately
$1/3$ for the $s^{0.0808}$ coefficients.
The total $J/\Psi-p$ cross section is $4$~mb.
The ratio of cross sections depends on the propagators and form factors.
We used the following mean squared radii~\cite{pov}:
$\langle{r_p^2}\rangle=0.67\rm~fm^2$, $\langle{r_{\pi}^2}\rangle=0.44\rm~fm^2$,
$\langle{r_K^2}\rangle=0.35\rm~fm^2$,
$\langle{r_{\Psi}^2}\rangle=0.04\rm~fm^2$, and
also from a non-relativistic quark model calculation $\langle
{r_{\Psi}^2}\rangle=0.06\rm~fm^2$.

The ratios $\sigma_{Kp}/ \sigma_{{\pi}p}$ and
$\sigma_{{\Psi}p}/\sigma_{{\pi}p}$ are shown in Fig.~4 as
a function of the gluon mass. The ratio $\sigma_{Kp}/\sigma_{{\pi}p}$
is pratically constant, and $\sigma_{{\Psi}p}/\sigma_{{\pi}p}$ exhibits an
appreciable variation with $m_g$. This can be understood since
for a perturbative propagator with zero gluon mass, and the form-factor in the
pole approximation
the cross section dependence on hadron radius, e.g.\ for the
collision of two identical mesons, is given by~\cite{dol}
%
$\sigma_M=\frac{64}{27}\pi\alpha_s^2\langle{r_M^2}\rangle$,
%
where $\langle{r_M^2}\rangle$ sets the scale of the cross
section. In the $J/\Psi$ case, the scale $\langle{r_M^2}\rangle$ is too small
creating a strong dependence of $\sigma_T$ and $m_g$. This result can be tested
experimentally.  The curves of Fig.~4
 are obtained for
$\Lambda=0.3$~GeV. With $m_g=0.37$~GeV we have
$\sigma_{Kp}/\sigma_{{\pi}p}\simeq0.92$ (to be compared with 0.87), and
$\sigma_{{\Psi}p}/\sigma_{{\pi}p}\simeq0.29$ with $\langle{r_{\Psi}^2}\rangle
\simeq 0.04\rm~fm^2$.  We predict the Pomeron contribution to
$\sigma_{{\Psi}p}$ to be equal to~$3.95s^{0.0808}$.

\section{Conclusions.}

The NPG model describes the Pomeron as the exchange of two non-perturbative
gluons. At low $k^2$ the gluon propagator shows the presence of a
dynamically generated gluon mass, and at high $k^2$ has the correct QCD
asymptotic behavior. It was computed the elastic cross section
for $pp$ scattering, finding agreement with experiment for $m_g=0.37$~GeV, when
$\Lambda=0.3$~GeV, which is consistent with Cornwall's determination
of $m_g$ through the gluon condensate. Here we obtain the Pomeron coupling to
quarks, the exclusive $\rho$ production in DIS, the
ratio of total cross section of $J/\Psi$-$p$ to $\pi$-$p$ scattering, and
determined the behavior $3.95s^{0.0808}$ for the Pomeron contribution to
$J/\Psi$-$p$ scattering. The results are consistent with a gluon mass of
$0.37$~GeV when $\Lambda_{\rm QCD}=0.3$~GeV.

\section*{Acknowledgments}
I would like to acknowledge my collaborators F.Halzen and A.A.Natale
for introducing me to Pomeron matters and for the hospitality at Phenomenology
Institute at the University of Wisconsin.
This research was supported in part by the
University of Wisconsin Research Committee with funds granted by the Wisconsin
Alumni Research Foundation, by the U.~S.~Department of Energy under contract
No.~DE-AC02-76ER00881, by the Texas National Research Laboratory Commission
under Grant No.~RGFY93-221, and by the  Conselho Nacional de Desenvolvimento
Cientifico e Tecnologico, CNPq (Brazil).


\section*{Figure Captions}

\noindent
{\bf Fig.~1.} Pomeron coupling to quarks $(\beta_0^2)$
as a function of the gluon
mass $m_g$ for different values of $\Lambda$.

\noindent
{\bf Fig.~2.} Total cross section for $pp$ as a function of the gluon mass for
different
values of $\Lambda$.

\noindent
{\bf Fig.~3.} Total cross section for exclusive $\rho^{0}$ production as
a function of the gluon mass $m_g$. Data from Ref.~[7].

\noindent
{\bf Fig.~4.} Ratios of total cross sections:
$\sigma_{Kp}/\sigma_{{\pi}p}$
(dashed curve), $\sigma_{{\Psi}p}/\sigma_{{\pi}p}$
with $\langle{r_{\Psi}^2}\rangle=0.04\rm~fm^2$
(solid curve), and $\langle{r_{\Psi}^2}\rangle=0.06\rm~fm^2$
(dotted curve). The curves were determined for $\Lambda=0.3$~GeV.

\end{document}